\documentclass[twoside]{article}
\usepackage{qic}
\usepackage[lofdepth,lotdepth,caption=false]{subfig}
\usepackage{graphicx}
\usepackage{color}
\usepackage{amsmath}

\renewcommand{\thefootnote}{\fnsymbol{footnote}}

\begin{document}
\setlength{\textheight}{8.0truein} %FOR 2ND PAGE ONWARDS

%\runninghead{Leakage Suppression in the Toric Code}
%            {M. Suchara, A. W. Cross, J. Gambetta}

\normalsize\textlineskip
\thispagestyle{empty}
\setcounter{page}{1}

%\copyrightheading{Vol.}{No.}{Year}{Page Nos.}
%\copyrightheading{0}{0}{2003}{000--000}

\vspace*{0.00truein}

\alphfootnote

\fpage{1}

\centerline{\bf
LEAKAGE SUPPRESSION IN THE TORIC CODE}
%\vspace*{0.035truein}
%\centerline{\bf FOR THE TORIC CODE}
\vspace*{0.37truein}
\centerline{\footnotesize
%%%%%%%%%%%%%%%%%%
%put authors' name and address here
%%%%%%%%%%%%%%%%%%
MARTIN SUCHARA\footnote{Corresponding author: msuchar@us.ibm.com}, ANDREW W. CROSS, JAY M. GAMBETTA}
\vspace*{0.015truein}
\centerline{\footnotesize\it IBM T. J. Watson Research Center, 1101 Kitchawan Road}
\baselineskip=10pt
\centerline{\footnotesize\it Yorktown Heights, New York 10598,
U.S.A.}
\vspace*{0.225truein}
%\publisher{(received date)}{(revised date)}

\vspace*{0.21truein}

\abstracts{
Quantum codes excel at correcting local noise but fail to correct leakage faults that excite qubits to states outside the computational space. Aliferis and Terhal \cite{aliferis07} have shown that an accuracy threshold exists for leakage faults using gadgets called leakage reduction units (LRUs). However, these gadgets reduce the accuracy threshold and can increase overhead and experimental complexity, and these costs have not been thoroughly understood. Our work explores a variety of techniques for leakage-resilient, fault-tolerant error correction in the context of topological codes. Our contributions are threefold. First, we develop a leakage model that differs in critical details from earlier models. Second, we use Monte-Carlo simulations to survey several syndrome extraction circuits. Third, given the capability to perform three-outcome measurements, we present a dramatically improved syndrome processing algorithm. Our simulation results show that simple circuits with one extra CNOT per qubit and no additional ancillas reduce the accuracy threshold by less than a factor of $4$ when leakage and depolarizing noise rates are comparable. This becomes a factor of $2$ when the decoder uses 3-outcome measurements. Finally, when the physical error rate is less than $2\times 10^{-4}$, placing LRUs after every gate may achieve the lowest logical error rates of all of the circuits we considered. We expect the closely related planar and rotated codes to exhibit the same accuracy thresholds and that the ideas may generalize naturally to other topological codes.
}{}{}

\vspace*{10pt}

\vspace*{3pt}
%\communicate{to be filled by the Editorial}

\vspace*{1pt}\textlineskip %) USE THIS MEASUREMENT WHEN THERE IS
   %) A SECTION HEADING
%\vspace*{-0.5pt}
%\noindent
%%%%%%%%%%%%%%%%
%put the text of the paper here
%%%%%%%%%%%%%%%%
\setcounter{footnote}{0}
\renewcommand{\thefootnote}{\alph{footnote}}

\section{Introduction}

Large-scale quantum computers require a fault-tolerant architecture based on quantum error-correcting codes. While there are many approaches to quantum fault-tolerance, topological codes~\cite{kitaev03,bravyi98} stand out due to their many favorable properties, including local check operators, simple syndrome extraction circuits~\cite{dennis02}, and flexible fault-tolerant logic based on transversal gates, code deformation~\cite{rh07,raussendorf07}, or lattice surgery~\cite{horsman12}. These properties endow topological codes with a high accuracy threshold. The value of the surface code's accuracy threshold depends on assumptions regarding the noise model and error-correction approach, but estimates vary from $0.67\%$ \cite{stephens14} to above $1\%$ \cite{wang11}.

These threshold estimates assume a depolarizing noise model that approximates realistic noise, but the model does not include so-called leakage faults. Leakage faults map quantum states out of the 2-dimensional qubit subspace and into a higher-dimensional Hilbert space. Many physical realizations of qubits are multilevel systems that we expect to suffer from leakage faults. Leakage errors can be mitigated by constructing quantum circuits with suitable properties. Broadly speaking, such gadgets convert leakage errors into ``regular'' errors and may or may not simultaneously raise a flag to indicate the location of the leakage error. Gadgets that detect leakage convert it into a located loss error that is easier to correct~\cite{gottesman97,preskill98,knill98}. Leakage reduction units (LRUs) convert leakage errors into regular errors but do not give an indication that leakage has occurred~\cite{aliferis07, fong11}. Leakage mitigation with LRUs was rigorously analyzed by Aliferis and Terhal~\cite{aliferis07}, who showed existence of a threshold to leakage in concatenated codes. Subsequent work has shown that circuits can be further simplified while remaining fault-tolerant \cite{fortescue14}.

Most studies of quantum codes do not consider errors due to leakage, and the cost of leakage mitigation has not been thoroughly understood, particularly in the surface code. When leakage errors are converted to detected loss, topological codes are known to have a 50\% loss threshold in an idealized model~\cite{stace09,fujii12}. However, the strategy of~\cite{stace09} measures high weight ``super'' check operators. While this may be natural for measurement-based quantum computing \cite{barrett10}, it is more difficult with a fixed planar array of qubits, as is the case with some superconducting architectures \cite{chow14}. Furthermore, if leakage errors instead remain undetected, or are detected but could have originated from multiple space-time points in the circuit, they may lead to additional error spread. Leakage of this form has been studied for a quantum repetition code~\cite{fowler13} and for specific types of gates applied to superconducting qubits~\cite{ghosh13, ghosh14}. Despite this work, the impact of leakage on the accuracy threshold of topological codes remains unclear. To address this problem, we systematically study several leakage reducing circuits and develop a new syndrome processing strategy that further enhances the benefits of these circuits.

The paper is organized as follows. Section~\ref{sec:toriccode} reviews the toric code and fault-tolerant error-correction using bare stabilizer measurements. Section~\ref{sec:leakagemodel} introduces a new model of leakage faults that is amenable to simulation within the stabilizer formalism, much like the standard depolarizing noise model. Section~\ref{sec:reduction} reviews circuit-based leakage reduction techniques, introduces the leakage mitigation circuits we study and why they work, and explains how matching-based syndrome processing algorithms are modified for these circuits. Section~\ref{sec:heralded} then proposes a model where we are given 3-outcome projective measurements for detecting leaked qubits and shows how to further modify syndrome processing algorithms to take advantage of this extra information. Section~\ref{sec:simulation} explains our simulation approach and relevant details of the implementation. Finally, Section~\ref{sec:results} presents our numerical results and Section~\ref{sec:conclusion} concludes.

\section{Toric Code}\label{sec:toriccode}

\begin{figure}[t]
\vspace*{13pt}
\centering
 \subfloat[short for lof][]{
 \epsfig{file=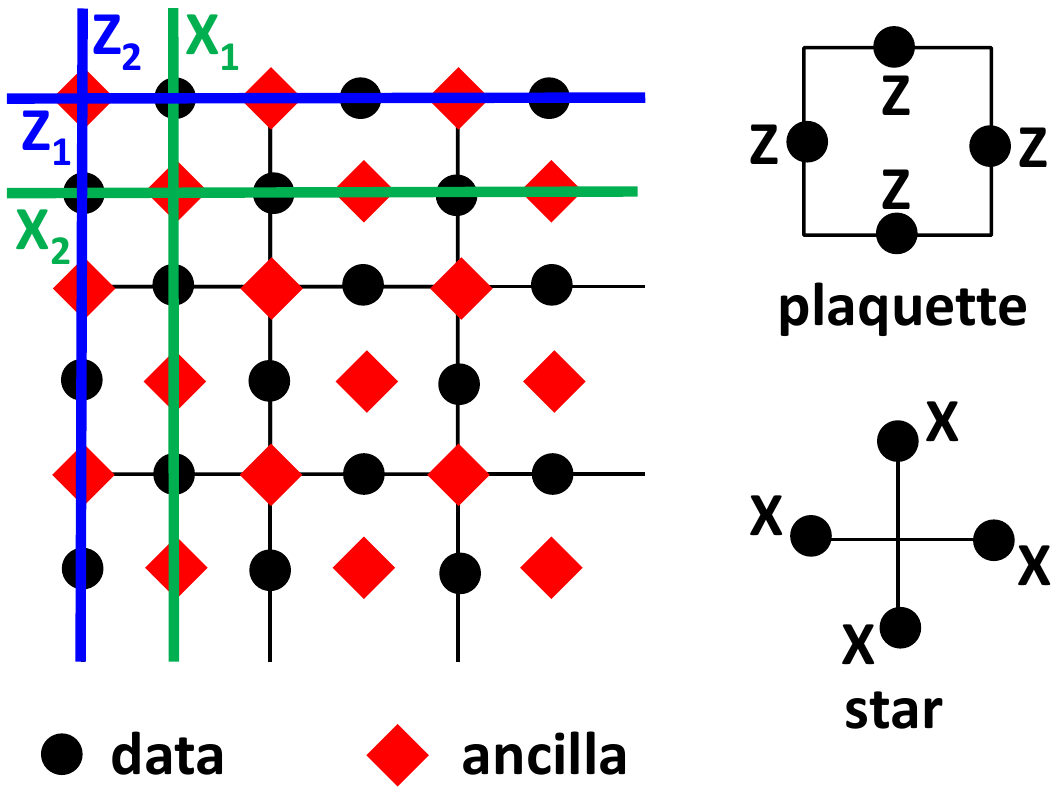,width=.31\textwidth}
   \label{subfig:lattice}
 }
 \subfloat[short for lof][]{
\epsfig{file=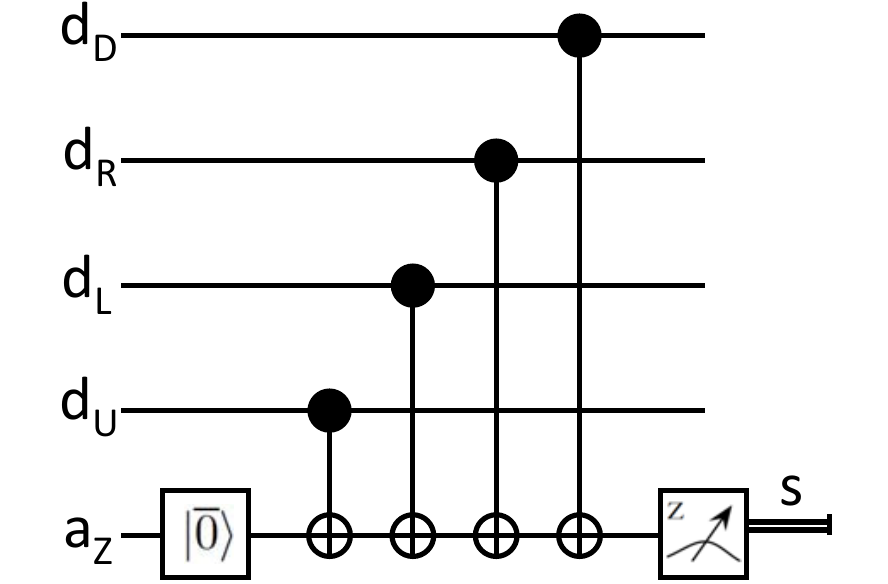,width=.33\textwidth}
   \label{subfig:ZZZZ}
}
 \subfloat[short for lof][]{
\epsfig{file=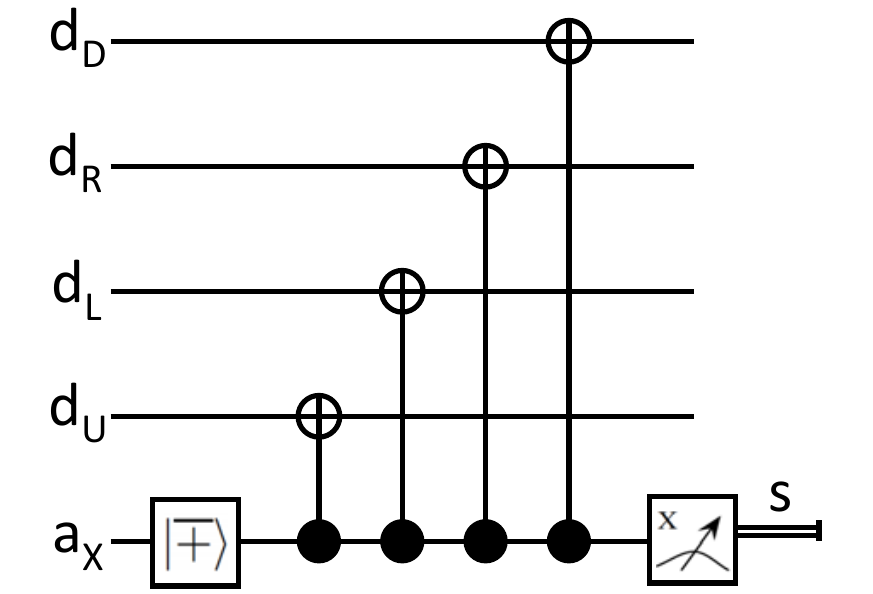,width=.33\textwidth}
   \label{subfig:XXXX}
 }
\vspace*{13pt}
\fcaption{\label{fig:toric} (Color online) (a) The toric code has a natural two-dimensional layout on the surface of a torus. Qubits function as either data qubits, used to store the encoded quantum state, or ancilla qubits, used to measure check operators (stabilizers) of the quantum code. Z-type check operators associate to faces (plaquettes) and X-type check operators associate to vertices (stars). Each check operator involves four data qubits and is measured using an ancilla qubit. The torus encodes a pair of qubits whose representative logical Pauli operators $X_1$, $Z_1$, $X_2$, and $Z_2$ are shown. (b,c) These circuits measure (b) plaquette and (c) star operators using four CNOT gates together with preparations and measurements. In each circuit, data qubits $d_{D,R,L,U}$ interact with an ancilla qubit $a_{z/x}$ that is prepared in the state $|0\rangle$ or $|+\rangle=\frac{1}{\sqrt{2}}\left(|0\rangle+|1\rangle\right)$. This ancilla is then measured in the basis of eigenstates of Z or X, respectively.}
\end{figure}

The toric codes \cite{kitaev03} are the prototypical example of topological stabilizer codes. Each toric code is defined on a $d$ by $d$ square array whose left-right and top-bottom boundaries are associated. The vertices of the array are connected to form a graph with $d^2$ vertices, $2d^2$ edges, and $d^2$ faces. Each of the $n=2d^2$ edges carries a physical qubit of the code, called a code or data qubit, and each vertex and face carries an ancillary qubit used for error-correction, called a syndrome or ancilla qubit (see Fig.~\ref{subfig:lattice}). The stabilizer $S$ of the quantum code is generated by a set of check operators $\{A_v\}$ and $\{B_f\}$ that belong to the $n$-qubit Pauli group and are attached to each vertex $v$ and face $f$ of the graph. These operators are tensor products of single qubit Pauli operators

\begin{equation}
X = \left(\begin{array}{cc}0&1\\1&0\end{array}\right)\ \textrm{and}\ Z = \left(\begin{array}{cc}1&0\\0&-1\end{array}\right).
\end{equation}
The vertex (or star) operators
\begin{equation}
A_v =\bigotimes_{\varepsilon\in N(v)}X_\varepsilon
\end{equation}
are $X$-type checks that apply a Pauli $X$ operator to the qubit on each edge $\varepsilon$ in the neighborhood $N(v)$ of vertex $v$. The neighborhood $N(v)$ of a vertex is the set of edges incident to $v$. The face (or plaquette) operators
\begin{equation}
B_f =\bigotimes_{\varepsilon\in N(f)}Z_\varepsilon
\end{equation}
are $Z$-type checks that apply a Pauli $Z$ operator to the qubit on each $\varepsilon$ in the neighborhood $N(f)$ of a face $f$. The neighborhood $N(f)$ of a face is the set of edges on the boundary of $f$. Each check operator involves only four code qubits since each vertex is incident to four edges and each face is bounded by four edges. The set of generators clearly commutes since any star shares zero or two edges with any face. Due to the periodic boundary conditions, any set of $d^2-1$ faces and $d^2-1$ vertices associate to an independent set of check operators that generate the stabilizer of the toric code. This implies that the toric code encodes a pair of logical qubits. Representatives for each class of logical Pauli $X$ and $Z$ operators are shown in Fig.~\ref{subfig:lattice}.

The overcomplete set of $2d^2$ check operators are measured simultaneously \cite{dennis02} using the circuits shown in Fig.~\ref{subfig:ZZZZ} and Fig.~\ref{subfig:XXXX}. In the first step, we prepare all plaquette ancillas in $|0\rangle$ and all site ancillas in $|+\rangle\propto |0\rangle+|1\rangle$. In the next four steps, CNOT gates act between each ancilla and the data qubit above, left, right, and below the ancilla, in that order. This corresponds to the gate order used in \cite{wang11,stephens14}. Finally, we measure each plaquette ancilla in the $Z$ eigenbasis and each star ancilla in the $X$ eigenbasis. These six steps constitute an error-correction cycle. One cycle produces a single noisy syndrome given by $2d^2$ bits.

A single error-correction cycle cannot be fault-tolerant in the toric code since local errors in the syndrome can lead to macroscopic errors after error-correction. However, $O(d)$ cycles suffice to improve confidence in the syndrome so that error-correction becomes fault-tolerant. A processing algorithm ingests these $O(d)$ syndromes and infers a corrective Pauli operator. The classic approach \cite{dennis02}, which we follow in this work, processes plaquette and site syndromes independently to find bit and phase error corrections. The problem of inferring the most probable error given the observed syndrome is mapped to a minimum weight perfect matching problem that can be solved with Edmond's algorithm \cite{edmonds65}.

\begin{figure} [b]
\vspace*{13pt}
\centerline{\epsfig{file=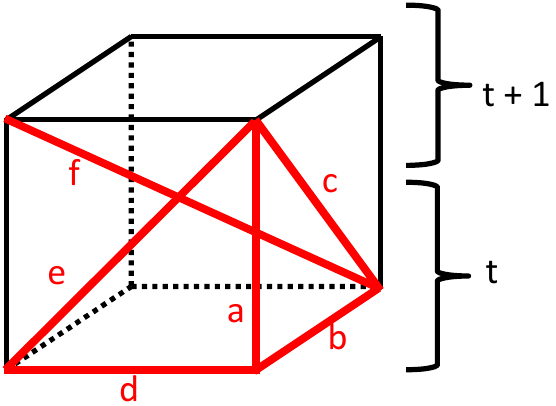, width=.30\textwidth}}
\vspace*{13pt}
\fcaption{\label{fig:cube} (Color online) The unit cell of the decoding graph has six distinctly-weighted edges shown here. Edges $b$ and $d$ correspond to qubit errors, edge $a$ to measurement error, edges $c$ and $e$ to correlated qubit-measurement errors, and edge $f$ to correlated qubit-qubit-measurement errors. The particular edges that appear are determined by the error-correction circuits.}
\end{figure}

The concept of a decoding graph ${\cal G}=({\cal V},{\cal E})$ is a useful abstraction for relating errors and corresponding syndromes \cite{bravyi13}. There is a separate decoding graph for bit flip ($X$) and phase flip ($Z$) errors. With the exception of Sec.~\ref{sec:heralded}, the graphs we consider are translation invariant, generated by translating the unit cell shown in Fig.~\ref{fig:cube} in each of three orthogonal directions. The edges ${\cal E}$ correspond to error events in the error-correction circuits, and, with the exception of diagonal edges $c$, $e$, and $f$ that correspond to correlated errors, are identified with qubits (horizontal edges ``$b$'' and ``$d$'') and measurement outcomes (vertical edges ``$a$'') for each error-correction cycle. An error on an edge $\varepsilon$ of ${\cal G}$ creates a pair of defects at the two vertices incident to $\varepsilon$. Errors on a subset of edges $E\subseteq {\cal E}$, called an error chain, create defects on $\partial E$ where $\partial E\subseteq {\cal V}$ is the set of vertices with an odd number of edges incident from $E$. Operationally, syndrome bit outcomes label the vertical ``$a$'' edges, and a defect occurs on vertex $v$ if the two incident ``$a$'' edges to $v$ have different syndrome labels.

The syndrome processing algorithm, or decoder, uses the minimum weight matching algorithm to match pairs of defects on the decoding graph. The error is then corrected by applying the $X$ or $Z$ correction on an error chain connecting each pair of matched defects. A chain $E$ is closed if $\partial E$ is empty, which occurs if the chain commutes with the stabilizer. A closed chain is contractible if the corresponding error belongs to the stabilizer. All error chains with the same $\partial E$ are equivalent up to closed chains, which include harmless contractible chains but also include non-contractible chains that are logical operators. The weight of each edge in the decoding graph is chosen to approximate the negative logarithm of the probability of the corresponding error. Therefore, independently for $X$ and $Z$ errors, a minimum weight matching decoder finds a correction from among the most likely errors and fails when the product of the correction and the error chain is not in the stabilizer.

In this work we restrict our attention to fault-tolerant quantum error-correction but our results apply to both active memory and computation \cite{rh07,horsman12}.

\section{Leakage Errors}\label{sec:leakagemodel}

Leakage errors are possible because the two-dimensional Hilbert space of a qubit is a subspace of a larger physical Hilbert space and interactions with the system may populate states in that larger physical space. The two-dimensional Hilbert space ${\cal H}_{C[j]}$ of the $j$th qubit is encoded into a Hilbert space ${\cal H}_{S[j]}={\cal H}_{C[j]}\oplus {\cal H}_{L[j]}$ where ${\cal H}_{L[j]}$ is an auxiliary leakage space associated to the qubit. We say that the $j$th qubit is {\em leaked} (resp. {\em contained}) if its state is supported entirely on ${\cal H}_{L[j]}$ (resp. ${\cal H}_{C[j]}$). The total Hilbert space of the system is ${\cal H}=\bigotimes_j{\cal H}_{S[j]}={\cal H}_S\oplus {\cal H}_L$ where ${\cal H}_S=\bigotimes_j{\cal H}_{C[j]}$ is the computational subspace and ${\cal H}_L$ is the remaining leakage subspace. States supported on ${\cal H}_L$ have at least one leaked qubit.

Consider a quantum computation implemented by one- and two-subsystem unitary gates that act on the total Hilbert space. For convenience we refer to these as one- and two-qubit gates. A {\em sealed single qubit gate} maps states in ${\cal H}_{C[j]}$ back to ${\cal H}_{C[j]}$ and therefore has the form $U_j = U_{{\cal H}_{C[j]}}\oplus U_{{\cal H}_{L[j]}}$, i.e. it does not leak. Similarly, a {\em sealed two-qubit gate} acting on states in ${\cal H}_{S[i]}\otimes {\cal H}_{S[j]}$ is a direct sum of unitary gates
\begin{equation}
U = U_{{\cal H}_{C[i]}\otimes {\cal H}_{C[j]}} \oplus
U_{{\cal H}_{C[i]}\otimes {\cal H}_{L[j]}} \oplus
U_{{\cal H}_{L[i]}\otimes {\cal H}_{C[j]}} \oplus
U_{{\cal H}_{L[i]}\otimes {\cal H}_{L[j]}}.\label{eq:idealgate}
\end{equation}
It is clear that the subspaces corresponding to each block are invariant under these gates. In this sense, sealed gates do not propagate leakage errors. In many physical settings, where gates are applied by a resonant or off-resonant drive or by tuning subsystems into resonance \cite{chow11,strauch13}, all blocks except ${\cal H}_{C[i]}\otimes {\cal H}_{C[j]}$ are approximately diagonal in the eigenbasis of the undriven system Hamiltonian and hence the physical gates are approximately sealed.

Leakage errors are generally coherent operations that are challenging to simulate for large quantum codes. Physical leakage noise may not be accurately captured by a Markovian noise model and generally needs to be treated as a coherent process \cite{aliferis07}. For simplicity and tractability, we define and simulate a stochastic leakage model, analogous to the standard approach of using depolarizing faults as a proxy for realistic decoherence and control faults. Although such a model does not accurately capture all conceivable leakage phenomena, it captures some aspects and yet remains amenable to stabilizer simulation.

Our stochastic model assumes ideal gates are sealed and that each subsystem is a 3-level system (a qutrit). We add three new elements to define faulty gates. The first two elements are rather straightforward. First, discrete leakage events occur independently with probability $p_\uparrow$ on each output qutrit of an ideal gate. These events are described by the map
\begin{equation}
{\cal E}_\uparrow(\rho_j)=(1-p_\uparrow)\rho_j + p_\uparrow |2\rangle\langle 2|
\end{equation}
where $\rho_j$ is the state of the $j$th qutrit. Second, there is a relaxation process with probability $p_\downarrow$, analogous to amplitude damping, that acts independently on each output qutrit of an ideal gate. The stochastic map for this process is given by
\begin{equation}
{\cal E}_\downarrow(\rho_j)=(1-p_\downarrow)\rho_j+p_\downarrow {\cal E}_{\mathrm{decay}}(\rho_j)
\end{equation}
where
\begin{equation}
{\cal E}_{\mathrm{decay}}(\rho_j)=A\rho_j A^\dagger + \sum_{k=0,1} A_{k}\rho A_{k}^\dagger
\end{equation}
is a decay map with elements
\begin{equation}
A = |0\rangle\langle 0| + |1\rangle\langle 1|,\ A_{k} = \frac{1}{\sqrt{2}}|k\rangle\langle 2|.
\end{equation}
These elements are chosen so that when a leaked qubit decays, it is replaced by a maximally mixed contained qubit. Let $Z^{(2)}=|0\rangle\langle 0|+|1\rangle\langle 1|-|2\rangle\langle 2|$ and suppose that an initial state $\rho_0$ is a fixed point of
\begin{equation}
\Lambda(\rho) = \frac{1}{2}\left( \rho + Z^{(2)}\rho Z^{(2)} \right).
\end{equation}
Then it is clear that any sequence of operations ${\cal E}_\uparrow$, ${\cal E}_\downarrow$, and sealed single qubit gates applied to $\rho_0$ produces a state that is also a fixed point of $\Lambda$, i.e. states are mixtures where qubits are leaked or contained. For convenience we call such mixtures on any number of qubits $|2\rangle$-{\em dephased}.

The final element of our model for faulty gates concerns two-qubit gates. Sealed two-qubit gates apply the intended operation to the ${\cal H}_{C[i]}\otimes {\cal H}_{C[j]}$ block and act arbitrarily on the remaining blocks. The ${\cal H}_{L[i]}\otimes {\cal H}_{L[j]}$ block consists of a single matrix element that applies what amounts to a harmless global phase in our model. The ${\cal H}_{C[i]}\otimes {\cal H}_{L[j]}$ and ${\cal H}_{L[i]}\otimes {\cal H}_{C[j]}$ blocks of a sealed gate merely apply a single qubit gate to qubit $i$ or $j$, respectively. As a final simplifying assumption, these blocks each implement Haar distributed random unitary gates, so that the resulting process completely depolarizes the contained qubit. Noisy two-qubit gates are followed by independent excitation and relaxation maps on each output qutrit. This is motivated by the assumption that matrix elements coupling computational states to $|22\rangle$ are significantly smaller than other matrix elements of the drive Hamiltonian. Like before, $|2\rangle$-dephased states remain so under the action of these two-qubit gates.

It is important to observe that this two-qubit gate model destroys correlations that could otherwise arise from interacting with leaked qubits. When multiple contained qubits interact with a leaked qubit, those contained qubits will experience independent depolarizing noise rather than some collective process. This would not be the case if, for example, the ${\cal H}_{C[i]}\otimes {\cal H}_{L[j]}$ and ${\cal H}_{L[i]}\otimes {\cal H}_{C[j]}$ blocks are fixed unknown single qubit unitaries. In this sense, our model is not a worst-case stochastic model. However, for topological codes, each leaked qubit interacts with a small set of neighbors and persists for at most a few syndrome extraction cycles before being reset by leakage reduction circuits. This would limit the impact of even a worst-case stochastic model but quantifying that impact we leave to future work.

Stabilizer circuits constructed from one- and two- qubit gates under the influence of our noise model can be simulated within the stabilizer formalism. The main observation is that $|2\rangle$-dephased states remain $|2\rangle$-dephased in our model, so we can introduce a classical {\em leakage indicator bit} for each qubit whose value indicates if the qubit is contained or leaked. We assume all qubits are initially contained. Quantum operations now act on an $n$-qubit stabilizer state and an $n$-bit leakage indicator array in the following way. The map ${\cal E}_\uparrow$ acts as identity with probability $1-p_\uparrow$ and otherwise traces out the corresponding qubit and sets the leakage indicator bit. Likewise ${\cal E}_{\mathrm{decay}}$ resets the leakage indicator bit and replaces the corresponding qubit, setting its state to be completely mixed. How sealed gates act on the $n$-qubit stabilizer state depends on the value of the leakage indicator array. Roughly speaking, the future quantum circuit is rewritten depending on the state of the leakage indicators. For our model, sealed one-qubit gates act as intended on contained qubits and as identity on leaked qubits. Likewise, sealed two-qubit gates also act as intended when both inputs are contained. However, a sealed two-qubit gate depolarizes one of the input qubits when the other input is leaked. When both inputs are leaked, a sealed two-qubit gate acts as identity. Taken together, and given a suitable definition of measurement for leaked qubits, one can sample from the output distribution of faulty stabilizer circuits built from these elements.

Often a full-blown stabilizer simulation is not needed for the study of fault-tolerant quantum error-correction and it suffices to track the Pauli error for each qubit. This works because we assume that the code qubits are always in an eigenstate of the stabilizer, and we can use the error operator to compute the eigenvalues of the stabilizers and normalizers and to predict deviations from ideal measurement outcomes without maintaining a description of the current quantum state as a set of stabilizer generators. For our leakage model, can we similarly divest ourselves of the full stabilizer framework and simply propagate for each qubit a Pauli error operator and a leakage indicator bit? The main concern is that the stabilizer circuit we simulate may not always measure operators in the current stabilizer when one or more leakage indicators are set. For example, leaked code qubits cause subsequent syndrome extraction circuits to measure check operators that are supported only on contained qubits. Likewise, leaked syndrome qubits ``turn off'' the associated check operator measurement. Without knowing the full stabilizer, we cannot predict how the measurement outcomes are correlated through parity constraints enforced by the stabilizer. The key observation that enables us merely to work with Pauli error and leakage indicator labels is that our model completely depolarizes syndrome qubits when code qubits are leaked, destroying any correlations that might otherwise have been present. Furthermore, when leaked qubits relax or pass through circuits to reset them, they are replaced by completely depolarized qubits. This guarantees that past syndrome history is uncorrelated with the particular state of the reset qubit (although it is correlated with the fact that the qubit has leaked). Reassured by this rough argument, we do not retain a complete set of stabilizer generators for the quantum state in the simulation. Instead, we simulate the model by maintaining and propagating a label ($I$, $X$, $Y$, $Z$ or $L$) for each qubit, where ``L'' denotes that the leakage indicator bit is set.

Now that we have explained our model and justified our simulation approach, here follows a summary of the details of the model. The model has parameters $p$ (depolarizing error probability), $q$ (syndrome measurement error probability, we set $q=p$), $p_\uparrow$ (probability of excitation outside of the computational state), and $p_\downarrow$ (probability of relaxation back to the computational state). For convenience, we also define the relative excitation rate $r = p_\uparrow / p$ and relaxation rate $s = p_\downarrow / p$. Elementary quantum gates behave as follows. Idle qubits are subject to depolarizing noise with probability $p$, where we apply one of $X$, $Y$, or $Z$ uniformly at random. They undergo free evolution and do not leak. However, a leaked qubit may relax to the computational space with probability $p_\downarrow$. The $|0\rangle$ or $|+\rangle$ state preparation succeeds with probability $1-p$, otherwise the orthogonal state is prepared. With probability $p_\uparrow$ the freshly prepared qubit leaks. If the input qubit is in the computational space, measurements report the incorrect outcome with probability $p$ and otherwise report the correct outcome. When the input qubit is leaked, we consider two scenarios: either the measurement cannot distinguish higher levels and reports ``1'', or the measurement has a third outcome ``L''. Noisy CNOT gates are subject to joint depolarizing noise with probability $p$, where we apply one of the non-identity two-qubit Pauli errors uniformly at random. In addition, if any input to the CNOT is leaked, all non-leaked qubits are completely depolarized, i.e. one of the $4$ single-qubit Pauli operators is applied uniformly to each non-leaked qubit. Finally, each output qubit leaks with probability $p_\uparrow$ and relaxes with probability $p_\downarrow$.

Unlike a similar model in~\cite{fowler13}, we assume that qubits reach a steady state leakage error rate at the beginning of any simulation, which more faithfully models the conditions of a long computation with the code. Consider the error-correction circuits in the standard toric code. Unlike ancillas, the data qubits are never re-initialized, and therefore they gradually accumulate leakage according to the transition probabilities $p_\uparrow$ and $p_\downarrow$ until an equilibrium is reached. It is easy to show from direct calculation of eigenstates of the transition matrix that the equilibrium distribution for data qubits in the toric code is given by $p_{eq}\approx\frac{4p_\uparrow}{4p_\uparrow+6p_\downarrow}$ where the factor $4$ is due to the four $CNOT$ gates, and the factor $6$ is due to the four $CNOT$s and two idle time steps. The second eigenvalue of the transition matrix is $(1-p_\downarrow)^6(1-p_\uparrow)^4$, so the non-equilibrium component of the distribution for each qubit decays as $\mathrm{exp}(-\zeta n)$ where $\zeta=-6\ln\left[(1-p_\downarrow)(1-p_\uparrow)\right]$. This suggests that starting from an equilibrium distribution may not be an overly pessimistic assumption. Under the reasonable assumptions $p_\uparrow \approx p_\downarrow$, the fraction of leaked data qubits reaches $40\%$ at which point the threshold is far exceeded and computation is clearly impossible (without leakage reduction, of course).

\section{Leakage Reduction}\label{sec:reduction}
\noindent

Leakage reduction is a process that converts leakage errors into regular errors that can be later corrected by quantum codes. Prompt leakage reduction is desirable as a single leaked qubit can cause many other errors (e.g., a leaked data qubit can damage ancillas which then spread more errors to other data qubits). A circuit that converts leakage on a single qubit into a regular error is a leakage reduction unit (LRU). LRUs satisfy two properties \cite{aliferis07}: (a) if their input is in the computational space then the identity operation is performed, and (b) if their input is in the leakage space then their output is some state in the computational space. We use LRUs as building blocks to construct several possible leakage suppressing error-correction circuits in the toric code.

\begin{figure}[b]
\vspace*{13pt}
\centering
\epsfig{file=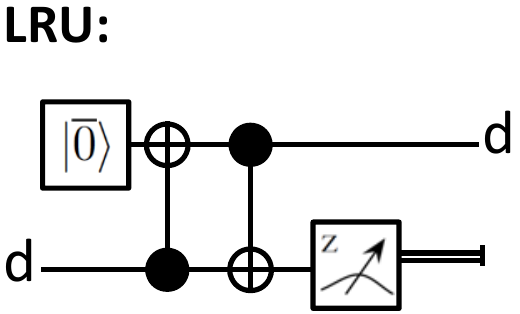,width=.19\textwidth}
\fcaption{\label{fig:lru} A leakage reduction unit (LRU) implemented by one-bit teleportation.}
\end{figure}

An LRU based on one-bit teleportation \cite{zhou00} is shown in Fig.~\ref{fig:lru}. This LRU differs from canonical LRUs based on full quantum teleportation \cite{mochon04,aliferis07}. It does not satisfy the properties of an LRU if ideal two-qubit gates can output a pair of leaked qubits when one of the inputs is leaked. In our model, ideal two-qubit gates are sealed and therefore the simpler one-bit teleportation circuit is an LRU and uses fewer resources than full quantum teleportation. Note that reinitializing qubits with most of their population in the computational space is essential to implementing any LRU in practice.

\begin{figure}[t]
\vspace*{13pt}
\centering
 \subfloat[short for lof][]{
 \epsfig{file=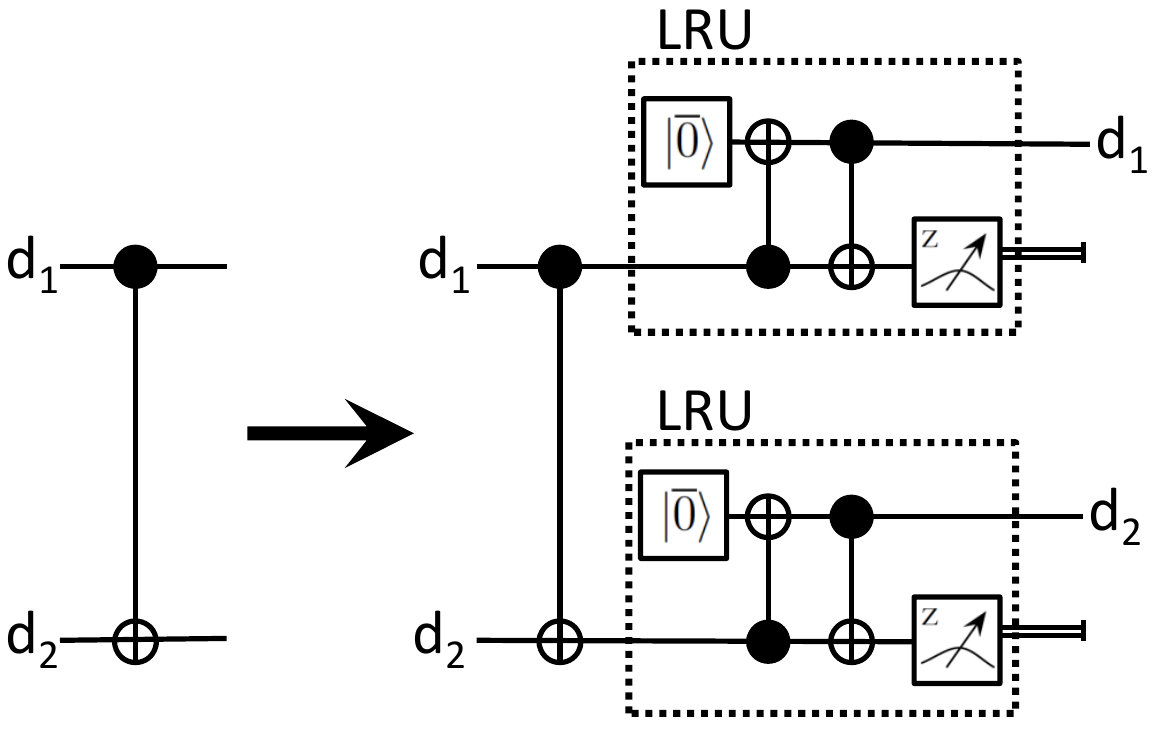,width=.37\textwidth}
   \label{subfig:full}
 }
 \subfloat[short for lof][]{
\epsfig{file=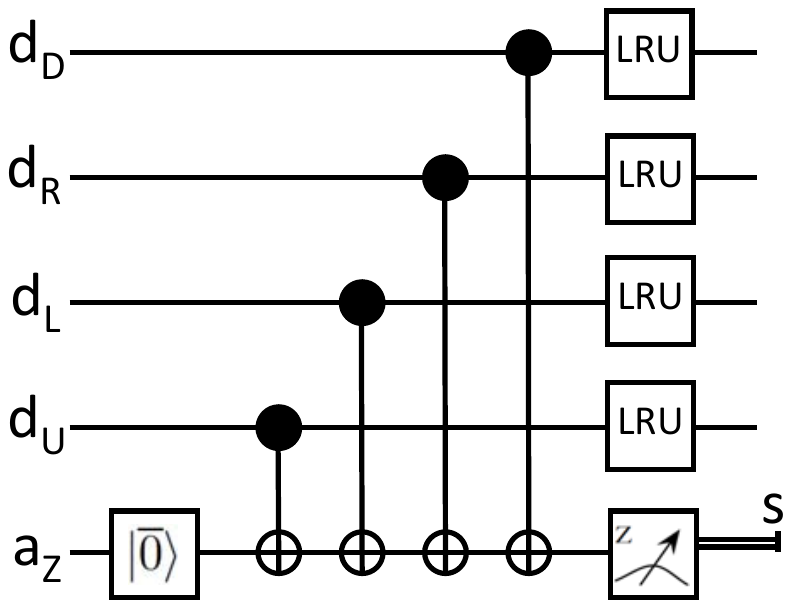,width=.27\textwidth}
   \label{subfig:partial}
}
 \subfloat[short for lof][]{
\epsfig{file=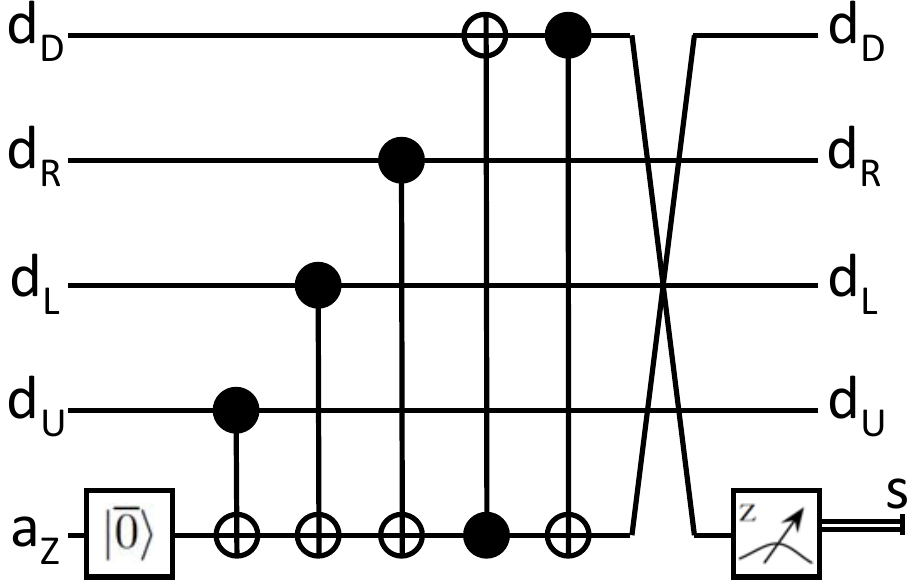,width=.32\textwidth}
   \label{subfig:quick}
 }
\vspace*{13pt}
\fcaption{\label{fig:LRcircuits} (a) The Full-LRU circuit replaces each gate with the gate followed by LRUs on each output qubit. (b) In the partial-LRU scheme, we insert an LRU on each data qubit at the end of each error-correction cycle. (c) The Quick scheme uses a circuit that is equivalent to those in Fig.~\ref{fig:toric}. The final CNOT gate acting on the ancilla and the $d_U$ data qubit has been replaced by a CNOT followed by a SWAP. This simplifies to a pair of CNOT gates.}
\end{figure}

We use this LRU to build several error-correction circuits that represent a tradeoff between the circuit complexity and effectiveness of leakage reduction. We list the circuits from the most complicated one with the best leakage suppression capability to the simplest one with the fewest extra ancillas:

\begin{enumerate}
\item \textbf{Full-LRU}: We call an error-correction circuit that replaces each gate by a rectangle containing the gate followed by LRUs on each output qubit (as in \cite{aliferis07}) the \emph{Full-LRU} circuit; see Fig~\ref{fig:LRcircuits}~(a). This circuit removes leakage immediately after each gate but uses a large number of additional gates and qubits per cycle. The Full-LRU circuit uses $16$ additional gates (in $4$ LRUs) per data and ancilla qubit per cycle. No LRU is applied to ancillas prior to measurement.

\item \textbf{Partial-LRU}: The \emph{Partial-LRU} error-correction circuit uses LRUs less frequently than the Full-LRU circuit. Once per error correction cycle, LRUs act on each data qubit while the ancillas are measured, as shown in Fig.~\ref{fig:LRcircuits}~(b). No additional effort is made for the ancillas, which are periodically measured and reinitialized anyway. While this strategy applies LRUs less frequently, which leads to greater spread of errors relative to the \emph{Full-LRU}, it only uses $4$ additional gates (in $1$ LRU) per data qubit per syndrome extraction and no additional gates applied to the ancillas.

\item \textbf{Quick}: The \emph{Quick} error-correction circuit further reduces the frequency of leakage reduction; see Fig.~\ref{fig:LRcircuits}~(c). At the end of each cycle, the circuit swaps each data qubit with an ancilla. A similar construction was proposed in~\cite{ghosh14} and shown to reduce leakage for a particular model relevant to superconducting qubits. In our case we choose to swap the ancilla with the data qubit $d_D$ immediately below the ancilla. Three $CNOT$ gates implement a $SWAP$ and, due to gate cancellation, there is only one additional $CNOT$ gate per data qubit compared to the standard circuit. After the swap, the physical qubits representing data and ancilla have traded roles. Each physical qubit is measured and reset every other cycle, so qubits do not remain leaked for many cycles.

\item \textbf{No LRU}: The last and simplest circuit we study is the standard syndrome extraction circuit depicted in Fig.~\ref{fig:toric}~(b,c), which we call \emph{No LRU} for convenience. The circuit is not suitable for error correction in the presence of leakage since data qubits are never reset. This circuit provides a point of reference from which to assess the effectiveness of the other circuits.
\end{enumerate}

While the Partial- and Full-LRU circuits are relatively straightforward to understand since they incorporate LRUs directly, the Quick circuit does not use LRUs. If there are no leakage errors, it is clear that the Quick circuit is functionally equivalent to the standard circuit. However, if an input ancilla or data qubit has leaked, one can calculate the effective circuit within our leakage model. Referring to Fig.~\ref{fig:LRcircuits}~(c), a leaked ancilla $a_z$ depolarizes each data qubit in the corresponding check, and the leaked qubit takes the role of $d_D$. The original $d_D$ is depolarized and measured, producing a corrupted syndrome bit. On the other hand, a leaked data qubit, which we take to be $d_D$, depolarizes the ancilla $a_z$. The data and ancilla exchange roles and the leaked qubit is measured. Meanwhile, $d_U$, $d_L$, and $d_R$ have interacted with $a_z$ before it was depolarized. One can check that these gates merely re-encode the plaquette operators and do not produce additional errors.

The realization of the Quick circuit in the toric code does not require extra ancillas. In the surface code, which does not have periodic boundary conditions, one could alternate swapping the ancillas with the $D$ and $U$ qubits in every odd and even error-correction cycle respectively. This approach only requires $O(d)$ additional qubits at the boundary of the lattice.

To decode syndromes produced by the leakage reducing circuits, we adjust the edge weights in the decoding graph so that the corresponding decoders again find a correction from among the most likely errors consistent with the syndrome. The decoder paired with the standard error-correction circuit (our No LRU circuit) uses a decoding graph depicted in Fig.~\ref{fig:cube}. The prior probability of an edge $\varepsilon$ (restricted to $\{a,b,c,d,e,f\}$ due to translation invariance) is taken as $p(\varepsilon)=\sum_j p_j(\varepsilon)$ where $p_j(\varepsilon)$ is the probability that a fault occurs at location $j$ and yields the syndrome $\partial\{\varepsilon\}$. The sum is taken over all locations in the circuit although not all locations contribute positive probability. If the physical error rates are sufficiently small, $p(\varepsilon)$ is approximately proportional to the probability that $\varepsilon$ carries an error. The edge weights are then given by $-\log p(\varepsilon)$. Edge weights for the standard circuit were approximated in this way in~\cite{wang11} by counting single-location faults, and we use an analogous approach to calculate the participating edges of the decoding graph and their weights for the other circuits introduced in this section. Specifically, we iterate over regular faults placed at single fault locations $j$, compute the corresponding syndrome and associated pair of defects, identify an edge $\varepsilon$ with the same pair of defects $\partial\{\varepsilon\}$, and add the probability of the fault $p_j(\varepsilon)$ to $p(\varepsilon)$.

\begin{table}[t]
\tcaption{Edges weights for each leakage reduction circuit}
\centerline{\footnotesize\smalllineskip
\begin{tabular}{r c c c c c c}\\
\hline
{} & $\mathrm{exp}(-w_a)$ & $\mathrm{exp}(-w_b)$ & $\mathrm{exp}(-w_c)$ & $\mathrm{exp}(-w_d)$ & $\mathrm{exp}(-w_e)$ & $\mathrm{exp}(-w_f)$ \\
\hline
{No LRU} & $31/15p+q$ & $28/15p$ & $16/15p$ & $52/15p$ & $8/15p$ & $8/15p$ \\
{Quick} & $7/3p+q$ & $32/15p$ & $4/3p$ & $4p$ & $8/15p$ & $8/15p$ \\
{Full-LRU} & $103/15p+q$ & $52/15p$ & $88/15p$ & $172/15p$ & $32/15p$ & $32/15p$ \\
{Partial-LRU} & $31/15p+q$ & $52/15p$ & $16/15p$ & $76/15p$ & $8/15p$ & $8/15p$ \\
\hline\\
\end{tabular}}
\label{table:weights}
\end{table}

The calculated edge weights in the decoding graph are summarized in Table~\ref{table:weights}. Here $p$ is the probability of gate, initialization, and idle error and $q$ is the probability of measurement error, which we set equal to $p$. The edge labels correspond to labels shown in Fig.~\ref{fig:cube}. The symmetry between the weights of edge pairs $b$-$d$ and $c$-$e$ is broken because the stabilizer measurements are not atomic operations but involve $CNOT$ gates applied in a particular order (see \cite{wang11}). Although the leakage reducing circuits differ from the No LRU circuit, there are no new edges in the decoding graphs of these circuits. One can see that, for example, the weights for the Full-LRU circuit on edges $e$ and $f$ are $4$ times the corresponding weights of the other circuits due to the $4$ locations in the LRU. The edge weights for the No LRU circuit agree with the weights in~\cite{wang11} except for edge $a$. Our probability for edge $a$ is higher because our error-correction circuits include ancilla initializations. We call a decoder that uses these edge weights the \emph{Standard decoder}. Note that the Standard decoder does not account for leakage events, and hence we expect it will be suboptimal.

\section{Heralded Leakage Reduction}\label{sec:heralded}
\noindent

Up to this point in the discussion, we have assumed that measurements cannot distinguish a leaked qubit from a qubit in the state $|1\rangle$, and therefore there was no mechanism to detect leakage faults. We now assume that measurements produce a third outcome ``L'' given a leaked qubit and optimistically assume that they output ``L'' if and only if the input qubit is leaked. Taking the same circuits considered in Sec.~\ref{sec:reduction} (specifically the Partial-LRU and Quick circuits), we modify the decoders to use this additional information. This approach is relatively simple, requiring no change to the quantum circuits, and leads to significantly improved logical error rates and threshold (see Sec.~\ref{sec:results}). A 3-outcome measurement is not equivalent to a simple erasure model because the precise space-time location of the leakage fault is not available, and leakage faults induce regular errors that depend on the leakage fault location.

The \emph{Heralded Leakage (HL) decoder} ingests results from 3-outcome measurements. Using these 3-outcome measurements provides an advantage because in our model leakage is likely to produce regular errors on specific locations around the leaked qubit, and the decoder is then able to match defects using correct prior probabilities in the space-time neighborhood of the ``L'' event in the decoding graph. Such events do not indicate exactly which gate suffered a leakage fault, since ``L'' events can only be detected by measuring a qubit, which only happens in an LRU or when measuring a syndrome. Therefore, upon registering an ``L'' event, the decoder only ``knows'' that the measured qubit leaked at some location between its initialization and measurement.

Conditioned on the observed ``L'' events, the HL decoder generates a new decoding graph from the Standard decoding graph ${\cal G}$ whose edge weights are given by Table~\ref{table:weights}. The ``L'' events are independent in our model, so suppose that a set of these events ${\cal L}$ occurs. Because our leakage model independently depolarizes qubits when they interact with a leaked qubit, we can construct a conditional decoding graph ${\cal G}_{\cal L}$, for bit-flips say, whose edges $\varepsilon$ again correspond to independent faults and whose edge weights are computed from conditional probabilities $p(\varepsilon|{\cal L})$. These conditional probabilities are computed in the same way as explained in Section~\ref{sec:reduction}. ${\cal G}_{\cal L}$ is generally not translation invariant and contains additional edges not present in Fig.~\ref{fig:cube}. There will be edges in ${\cal G}_{\cal L}$ with very low weight associated to highly probable errors on all of the qubits that interacted with leaked qubits.

\begin{figure}[b!]
\vspace*{13pt}
\centering
\begin{minipage}[c]{\textwidth}
\centering
\epsfig{file=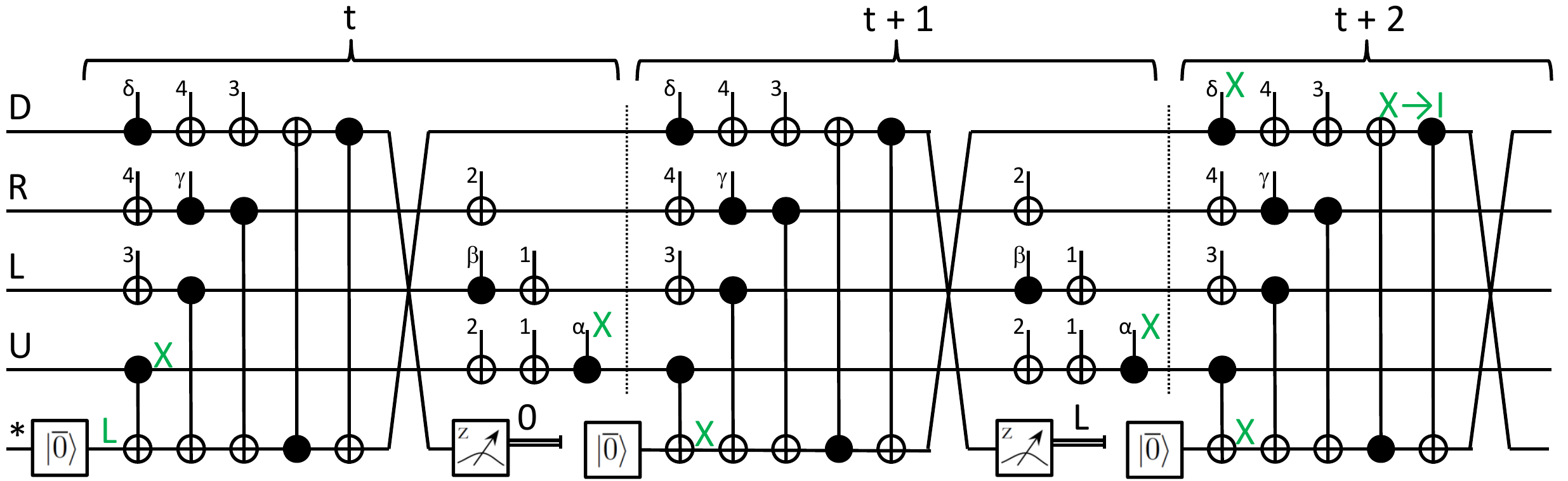, width=.88\textwidth}
\vspace*{13pt}
\fcaption{\label{fig:quickCircuit} (Color online) A measurement detects a leakage event in the Quick circuit at time $t+1$. This circuit depicts all of the gates in error-correction cycles $t$, $t+1$, and $t+2$. Each qubit that interacts with the potentially leaked qubit may be depolarized; subsequent gates in the cycle potentially spread the errors. The error labels show an example fault path (corresponding to the edge labeled $p_1$ in Fig.~6 (b)) where an ancilla leaks and introduces an $X$ error on qubit $U$ that propagates to other qubits. The notation ``$X\mapsto I$'' indicates a cancellation of errors.}
\vspace*{13pt}
\end{minipage}
\begin{minipage}[c]{\textwidth}
 \subfloat[short for lof][]{
 \epsfig{file=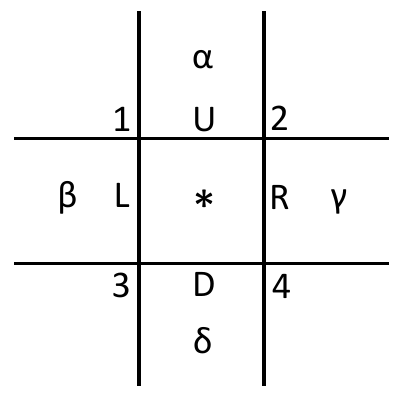,width=.23\textwidth}
   \label{subfig:quickKey}
 }
 \subfloat[short for lof][]{
\epsfig{file=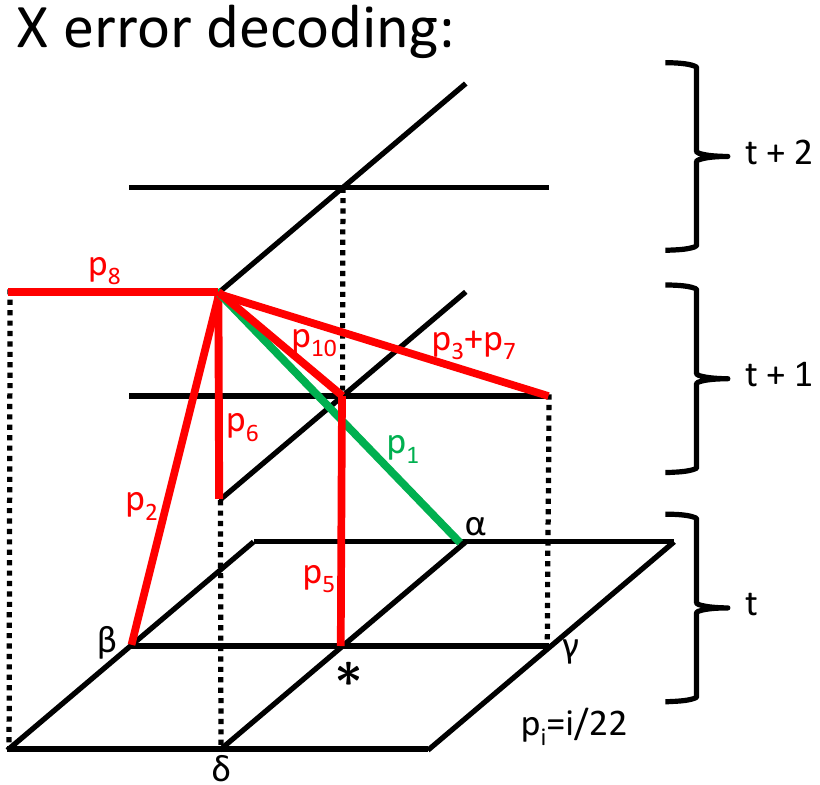,width=.36\textwidth}
   \label{subfig:quickXCube}
}
 \subfloat[short for lof][]{
\epsfig{file=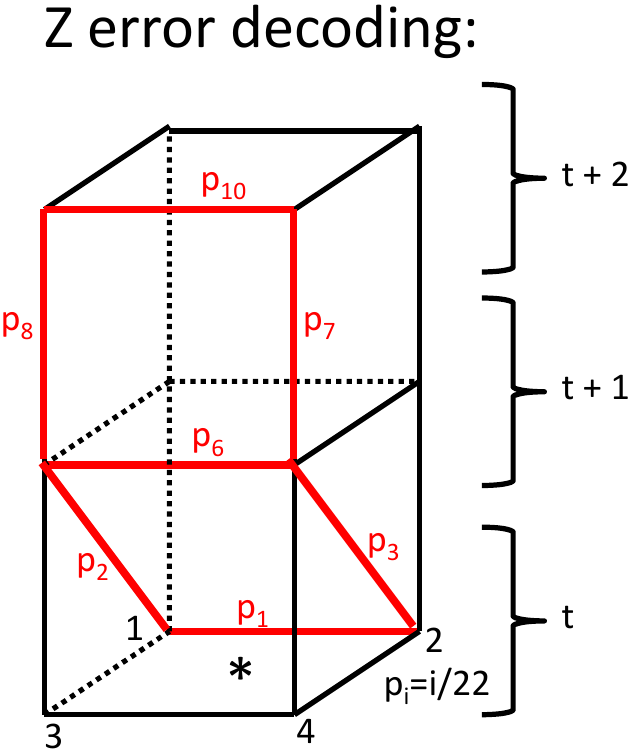,width=.27\textwidth}
   \label{subfig:quickZCube}
 }
\vspace*{13pt}
\fcaption{\label{fig:quick} (Color online) (a) The location of the detected leakage (labeled by $*$) and the relative locations of other qubits in the Quick circuit. (b,c) Upon leakage detection at time $t$ on the plaquette ancilla *, the edge weights in the decoding graphs are updated as indicated.}
\vspace*{13pt}
\end{minipage}
\end{figure}

If an ``L'' event occurs when measuring a syndrome, the corresponding syndrome bit is absent from the syndrome history and the usual approach for computing defect locations is somewhat modified. Suppose that ``L'' events occur for some check operator at $m$ contiguous times $t+1$, $\dots$, $t+m$. In this case, each of the corresponding ``$a$'' edges $e_{t+1}$, $\dots$, $e_{t+m}$ in the decoding graph are assigned weight $0$. Let $s_t$ be value of the check operator's syndrome bit at time $t$. If $s_t$ differs from $s_{t+m+1}$, we place a defect on the earliest vertex in $\partial\{e_{t+1}\}$.

To compute the edge weights in the conditional decoding graph, we proceed as follows, examining each ``L'' event independently. When an ``L'' event occurs on some qubit $q$, we consider each of the $n$ fault locations between and including $q$'s initialization and measurement. Since each gate suffers leakage faults independently and with equal probability, the probability that $q$ is leaked when interacting at the $i$-th location is approximately $i/n$. An $X$ and $Z$ error each occurs on the qubit that interacts with $q$ at location $i$ with probability $p_i = i/2n$. We consider each of these errors separately, find the defect pair they cause, and modify the weight of the edge connecting this defect pair. Since $p_i$ can be large, we take any probability $p_0$ previously associated to the edge and set the new probability equal to $p_0+p_1-2p_0p_i$.

\begin{figure}[b!]
\vspace*{13pt}
\centering
\begin{minipage}[c]{\textwidth}
\centering
\epsfig{file=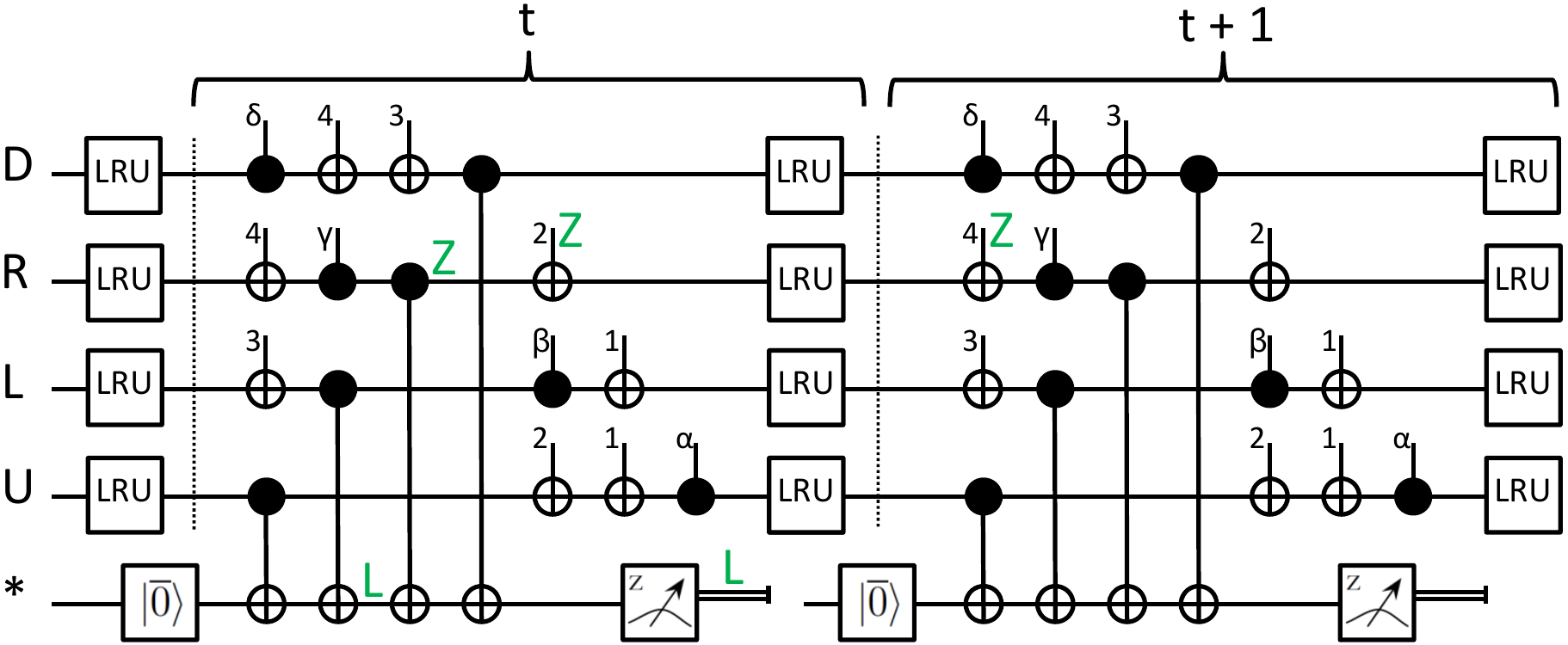, width=.67\textwidth}
\vspace*{13pt}
\fcaption{\label{fig:circuitAncillaCircuit} (Color online) A plaquette ancilla measurement detects a leakage event at time $t$ in the \emph{Partial-LRU} circuit. This circuit depicts all of the gates on this set of qubits in the $t$th and $t+1$th error-correction cycle. Each data qubit that interacts with the ancilla may be depolarized, and subsequent gates in the cycle potentially spread the errors. The error labels show an example fault path (corresponding to the edge labeled $p_3$ in Fig.~8 (c)) where an ancilla leaks and introduces a $Z$ error on qubit $R$ that propagates to other qubits.}
\vspace*{13pt}
\end{minipage}
\begin{minipage}[c]{\textwidth}
 \subfloat[short for lof][]{
 \epsfig{file=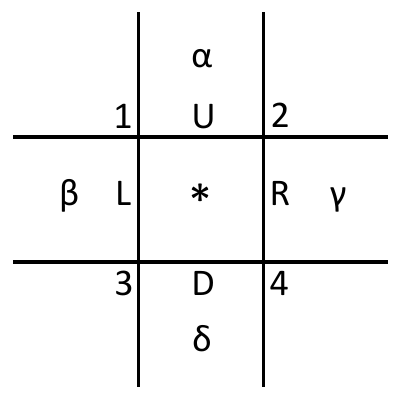,width=.23\textwidth}
   \label{subfig:circuitAncillaKey}
 }
 \subfloat[short for lof][]{
\epsfig{file=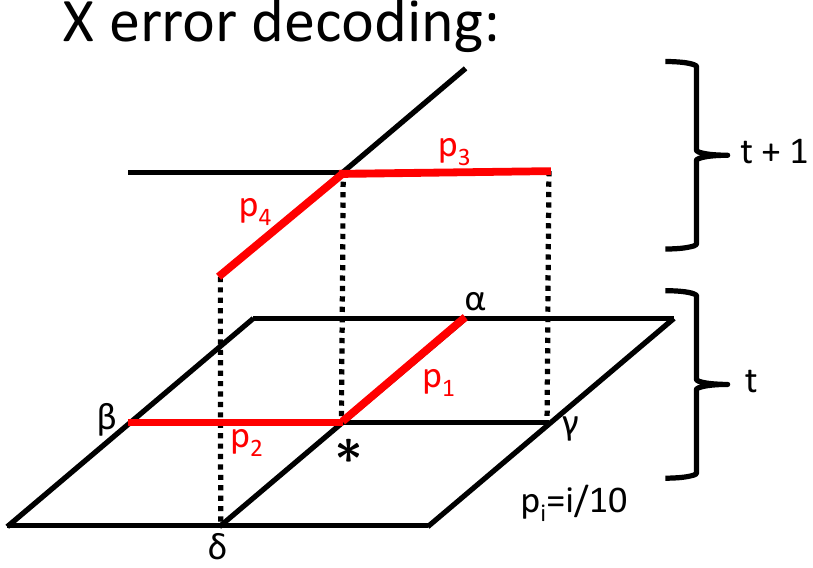,width=.36\textwidth}
   \label{subfig:circuitAncillaXCube}
}
 \subfloat[short for lof][]{
\epsfig{file=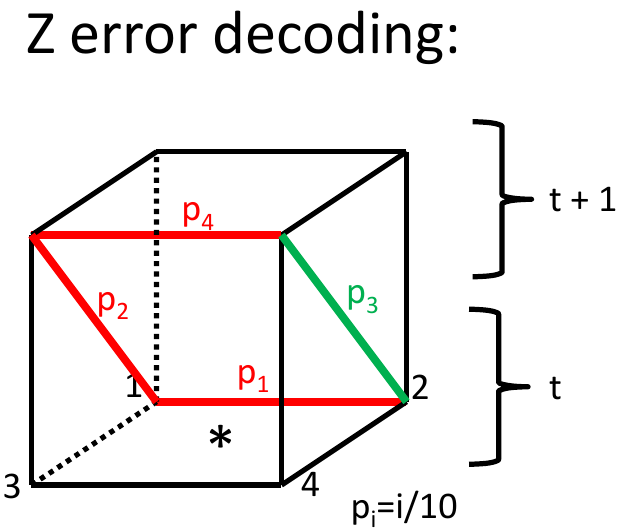,width=.27\textwidth}
   \label{subfig:circuitAncillaZCube}
 }
\vspace*{13pt}
\fcaption{\label{fig:circuitAncilla} (Color online) (a) The location of the detected leakage (labeled by $*$) and the relative locations of other qubits in the Partial-LRU scheme. (bc) Upon leakage detection at time $t$ on the plaquette ancilla *, the edge weights in the decoding graphs are updated as indicated.}
\vspace*{13pt}
\end{minipage}
\end{figure}

The added edges and their weights are obviously specific to each leakage reduction circuit, so we now describe how to construct conditional decoding graphs in greater detail for the Quick and Partial-LRU circuits.

Fig.~\ref{fig:quickCircuit} shows three rounds of plaquette measurements using the Quick circuit. Leakage is detected at time $t+1$ on the ancilla qubit labeled with $*$. Fig.~\ref{fig:quick} shows the relative position of the qubits in the circuit, and the low weight edges that are added to or modified in the $X$- and $Z$-error conditional decoding graphs. The edges are shown relative to the position of the leaked qubit.  Consider the following example in which we show that the long edge labelled $p_1$ in Fig.~\ref{subfig:quickXCube} is associated to an event that occurs with probability $\approx 1/22$ conditioned on the ``L'' event shown in Fig.~\ref{fig:quickCircuit}. This edge is not present in the standard decoding graph and is added to the conditional decoding graph for $X$ errors. The error labels in Fig.~\ref{fig:quickCircuit} show the events responsible for addition of the edge. A total of $11$ gates act on the leaked qubit between its initialization and measurement, and the qubit is leaked directly after initialization with probability $1/11$. Therefore, an $X$ error spreads to qubit $U$ at time $t$ with probability $1/2 \times 1/11$. This $X$ error spreads to qubit $\alpha$ at time $t$ and continues to be detected by the $\alpha$ plaquette from time $t$ onward. An $X$ error also spreads to the ancilla that replaces the ancilla $*$ at time $t+1$, which in turn spreads to qubit $\delta$ starting at time $t+2$, as indicated. Therefore, this error introduces the long edge labeled $p_1$ connecting the $\alpha$ and $\delta$ qubits in the $X$ error decoding graph. Other edges in the figure are obtained by considering other errors.

\begin{figure}[b!]
\vspace*{13pt}
\centering
\begin{minipage}[c]{\textwidth}
\centering
\epsfig{file=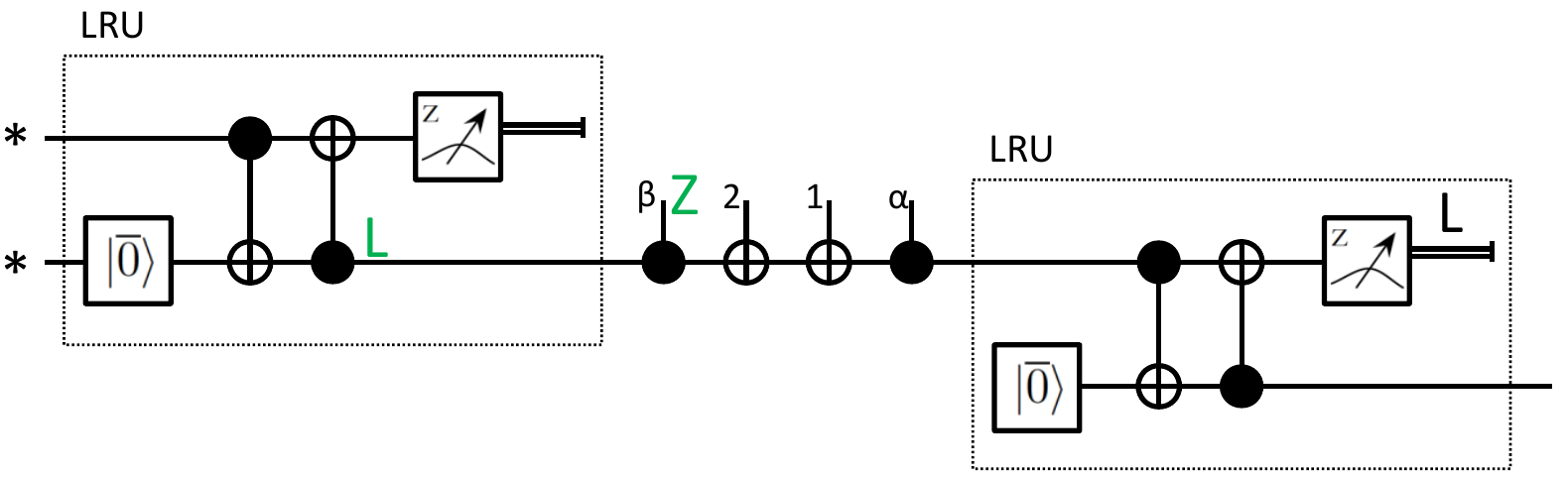, width=.61\textwidth}
\vspace*{13pt}
\fcaption{\label{fig:circuitHDataCircuit} (Color online) A data qubit's trailing LRU detects a leakage event at time $t$. This circuit depicts all of the gates that act on the data qubit in the $t$th error-correction cycle, together with the preceding LRU. Each ancillary qubit that interacts with the data may be depolarized, as well as the outgoing qubit of the trailing LRU, and subsequent gates in the cycle potentially spread the errors. The error labels show an example fault path (corresponding to the edge labeled $P_3$ in Fig.~10 (c)) where the data leaks and introduces a $Z$ error on an ancilla qubit.}
\vspace*{13pt}
\end{minipage}
\begin{minipage}[c]{\textwidth}
 \subfloat[short for lof][]{
 \epsfig{file=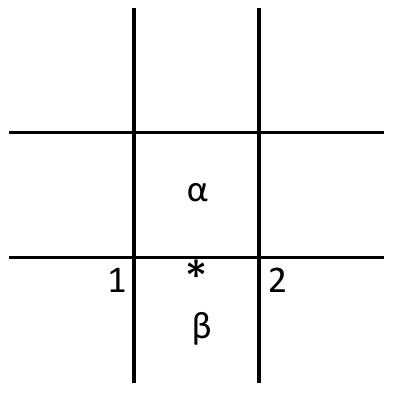,width=.23\textwidth}
   \label{subfig:circuitHDataKey}
 }
 \subfloat[short for lof][]{
\epsfig{file=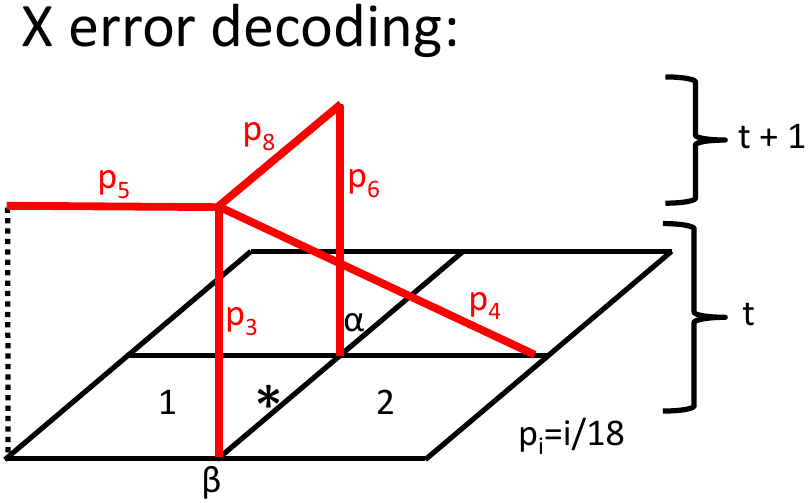,width=.36\textwidth}
   \label{subfig:circuitHDataXCube}
}
 \subfloat[short for lof][]{
\epsfig{file=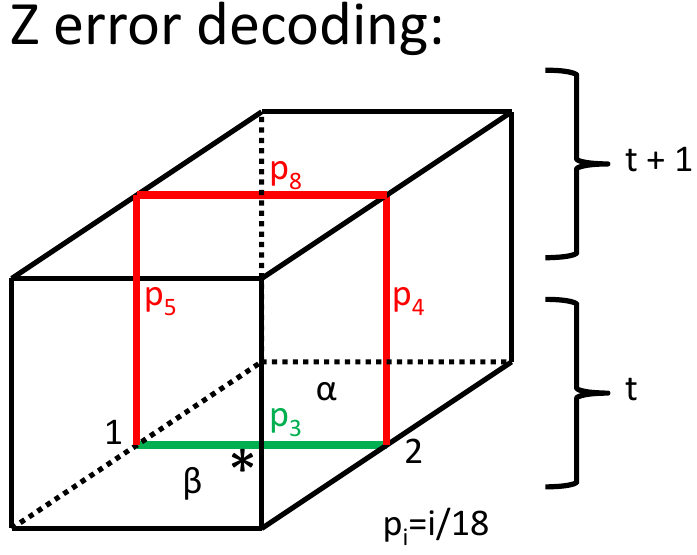,width=.30\textwidth}
   \label{subfig:circuitHDataZCube}
 }
\vspace*{13pt}
\fcaption{\label{fig:circuitHData} (Color online) (a) The location of the detected leakage (labeled by $*$) and the relative locations of other qubits in the Partial-LRU scheme. (bc) Upon leakage detection at time $t$ on the data qubit *, the edge weights in the decoding graphs are updated as indicated.}
\vspace*{13pt}
\end{minipage}
\end{figure}

Unlike the Quick circuit, the Partial-LRU circuit can detect leakage on ancillas, which are directly measured, and data qubits, which are measured by LRUs. First consider the ancillas. Two rounds of plaquette measurements are shown in Fig.~\ref{fig:circuitAncillaCircuit} with an ``L'' event at the ancilla $*$ at time $t$. The modified edges of the decoding graph, conditioned on the ``L'' event, are shown in Fig.~\ref{fig:circuitAncilla}. The error labels in Fig.~\ref{fig:circuitAncillaCircuit} show the events responsible for addition of the edge labeled $p_3$ in Fig.~\ref{subfig:circuitAncillaZCube}. The probability that the ancilla is leaked after the second $CNOT$ gate is $3/5$ and a $Z$ error spreads to qubit $R$ with probability $p_3=3/10$. The $Z$ error spreads to the qubits labeled $2$ and $4$ at time $t$ and $t+1$, and hence we modify the weight of the diagonal edge labeled $p_3$ connecting these two qubits in the decoding graph. Next, consider Fig.~\ref{fig:circuitHDataCircuit}, which shows LRUs acting on a data qubit. If the second LRU detects leakage, the edges shown in Fig.~\ref{fig:circuitHData} are added or modified to construct the conditional decoding graphs. The error labels in Fig.~\ref{fig:circuitHDataCircuit} correspond to the error responsible for addition of edge labeled $p_3$ in the $Z$ error decoding graph.

\section{Simulation Methods}\label{sec:simulation}
\noindent

We implemented simulations of the No LRU, Partial-LRU, Full-LRU, and Quick leakage reduction strategies. We evaluate both the Standard and HL decoders for the Partial-LRU and Quick strategies, and the Standard decoder for the No LRU and Full-LRU strategies. The simulation propagates labels $I$, $X$, $Y$, $Z$ or $L$ for each ancilla and data qubit in a toric code of distance $d$. We simulate the propagation of regular and leakage errors according to the error model described in Section~\ref{sec:leakagemodel}. Fault-tolerant error-correction in the toric code of distance $d$ requires $O(d)$ rounds of syndrome measurements so we simulate $d$ rounds. In addition, we include a round of perfect syndrome measurement at the end of each simulation where leaked qubits are replaced by completely depolarized qubits and the syndromes are measured without errors. All $d+1$ syndromes are processed together, a correction is found, and a failure is registered if the state of the code qubits after correction anticommutes with a logical operator of the code. Syndrome processing is performed such that the final state of the code qubits commutes with the stabilizer.

We assume that the $d$ rounds of fault-tolerant error-correction take place in the course of a longer computation. Therefore, each qubit is initially assigned the $L$ label with probability that is calculated based on the parameters of the model ($p_\uparrow$ and $p_\downarrow$) and on the frequency of the qubit re-initializations for the particular circuit. Specifically, persistent qubits that are never re-initialized, such as data qubits in the No LRU circuit, start in the $L$ state with the equilibrium probability derived in Section~\ref{sec:leakagemodel}.

The Standard decoding graph is built using the unit cell in Fig.~\ref{fig:cube}. The translation invariance of the decoding graph allows us to calculate the distance of two defects directly from their $(x, y, t)$ coordinates where the $x$ and $y$ specify the location of the defect on the torus and $t$ is the location in the time dimension. The decoding graph of the HL decoder is not translation invariant due to the presence of the low weight edges introduced in Section~\ref{sec:heralded}. Since calculating distance of the defects based on their coordinates would be difficult, we build the entire decoding graph and use a shortest path algorithm to find the distance between defects.

The implementation uses the Boost~\cite{Boost} libraries to find shortest paths, and the Blossom V~\cite{kolmogorov09} library to perform minimum weight perfect matching. Our results were generated by Monte Carlo simulations which were repeated until we reached at least $10,000$ iterations and at least $1,000$ failures for each configuration. These simulations were run on an IBM Blue Gene/Q~\cite{BGTeam11} using about 30,000 CPU-hours.
\section{Results}\label{sec:results}
\noindent

\begin{figure}[b!]
\vspace*{13pt}
\centering
 \subfloat[short for lof][]{
 \epsfig{file=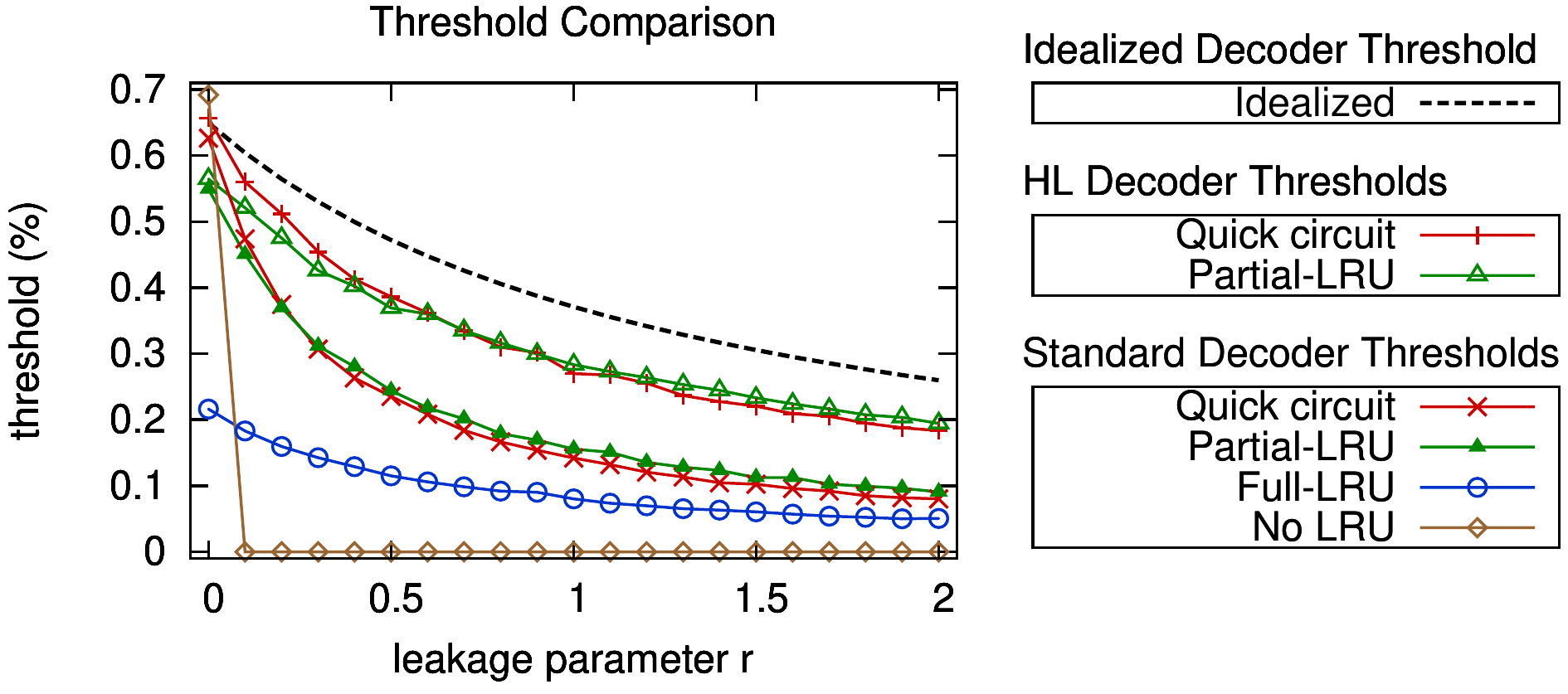,width=.595\textwidth}
   \label{subfig:all}
 }
 \subfloat[short for lof][]{
\epsfig{file=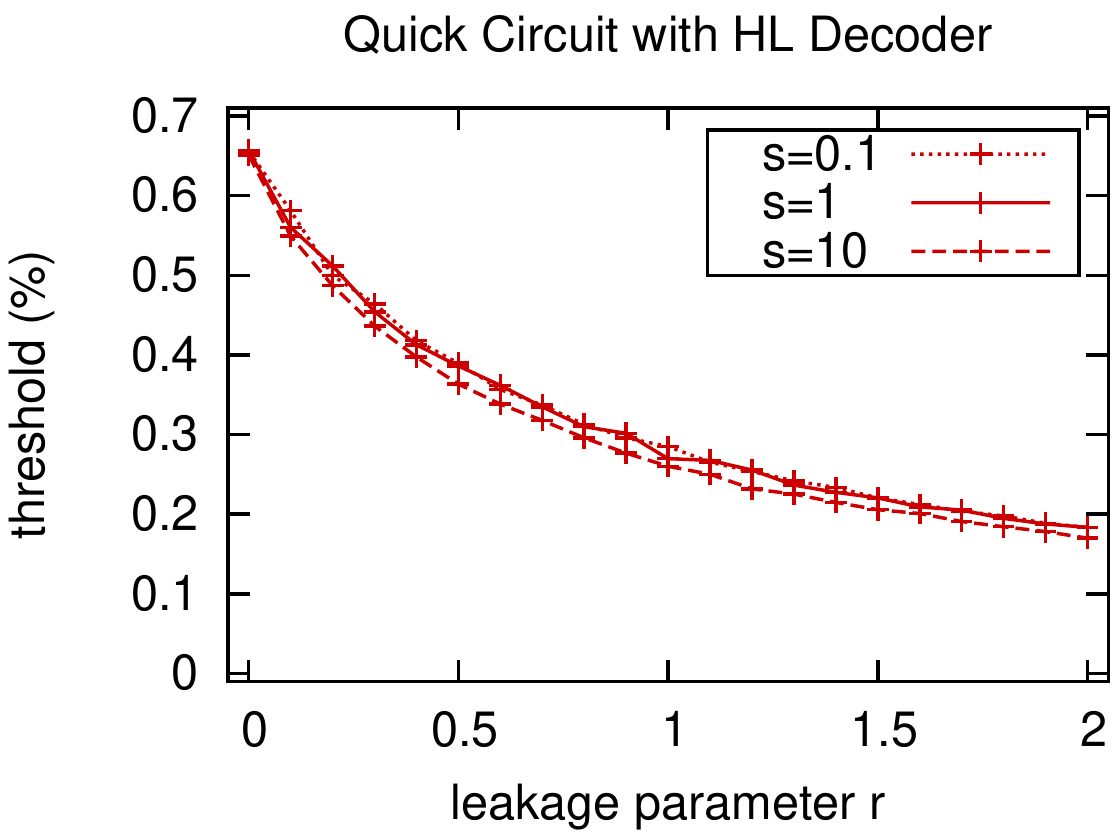,width=.365\textwidth}
   \label{subfig:pdexplore}
}
\vspace*{13pt}
\fcaption{\label{fig:thresholds} (Color online) (a) Summary of thresholds for the HL and Standard decoders. The HL decoder offers a substantially improved threshold. (b) The threshold, show here for the Quick circuit with HL decoder, does not change with the relative relaxation rate $s$.}
\end{figure}

A systematic exploration of the accuracy thresholds for the Standard and HL decoders is depicted in Fig.~\ref{fig:thresholds}~(a). The plot shows the accuracy threshold as a function of the amount of leakage (the relative excitation rate $r$). The plot was obtained by recording the failure rates for code distance $d=7$ and $d=9$ and by varying $r$ in increments of $0.1$ and $p$ in increments of $0.01\%$. We chose a relative relaxation rate $s = 1$. For each parameter choice we found the crossover of the failure rates for the two code distances, which approximates the threshold.

Based on Fig.~\ref{fig:thresholds}~(a) we make the following observations. In the regime with no leakage ($r = 0$), simpler circuits, i.e. circuit with fewer locations per syndrome extraction, have higher thresholds than more complicated circuits, in agreement with expectations. With no leakage, the No LRU circuit has a threshold of about $0.70\%$ whereas the Full-LRU circuit, which is the most complicated circuit we studied, has a threshold of about $0.22\%$. We also observe that the Quick and Partial-LRU circuits give an almost identical threshold, differing visibly only for $r<0.3$. The Quick circuit benefits from having fewer gates and qubits, whereas the Partial-LRU benefits from more frequent qubit reinitializations and therefore more effective leakage suppression. These features appear to have comparable effects on the threshold. The HL decoder significantly improves the threshold compared to the Standard decoder when $r$ exceeds about $0.3$.

As the $r$ increases, the threshold decreases monotonically for all circuits and decoders. To understand this, consider an idealized setting where we assume that leaked qubits are immediately replaced by completely depolarized qubits, causing errors $X$, $Y$, $Z$, and $I$ each with equal probability. This makes a leakage event equivalent to a regular depolarizing error with probability $\beta=3/4$. Suppose that physical errors occur at an effective rate $\tilde{p}=(1+\beta r)p$ that is the sum of two terms where the first one is due to regular depolarizing errors and the second one due to leakage. Let's also assume that there is some minimum number of faults $m$ that can produce a logical error. The logical error rate is then given by
\begin{equation}
p' = A\tilde{p}^m+O(\tilde{p}^{m+1}) = A(1+\beta r)^{m}p^m + O(p^{m+1}).
\end{equation}
Neglecting the higher order terms and solving for the fixed point gives us a very rough approximation for the threshold as
\begin{equation}
p_{th}^{(r)} \approx \left[A(1+\beta r)^m\right]^{-1/(m-1)}.
\end{equation}
We are interested in the ratio of thresholds when $r=0$ versus $r>0$, which is
\begin{equation}
\frac{p_{th}^{(r)}}{p_{th}^{(0)}} = (1+\beta r)^{-m/(m-1)} \approx \frac{1}{1+\beta r}.
\end{equation}

\begin{figure} [b!]
\vspace*{13pt}
\centering
 \subfloat[short for lof][]{
 \epsfig{file=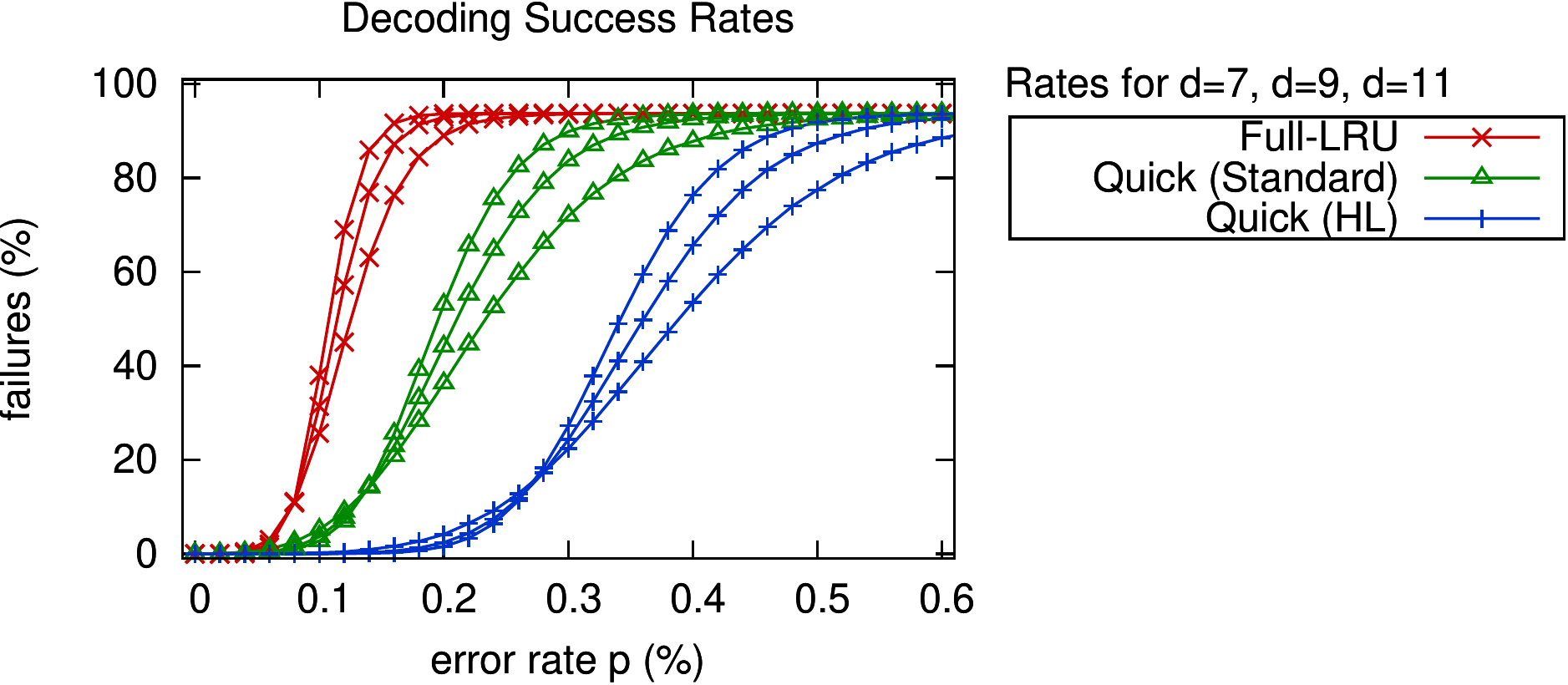,width=.585\textwidth}
   \label{subfig:linear}
 }
 \subfloat[short for lof][]{
\epsfig{file=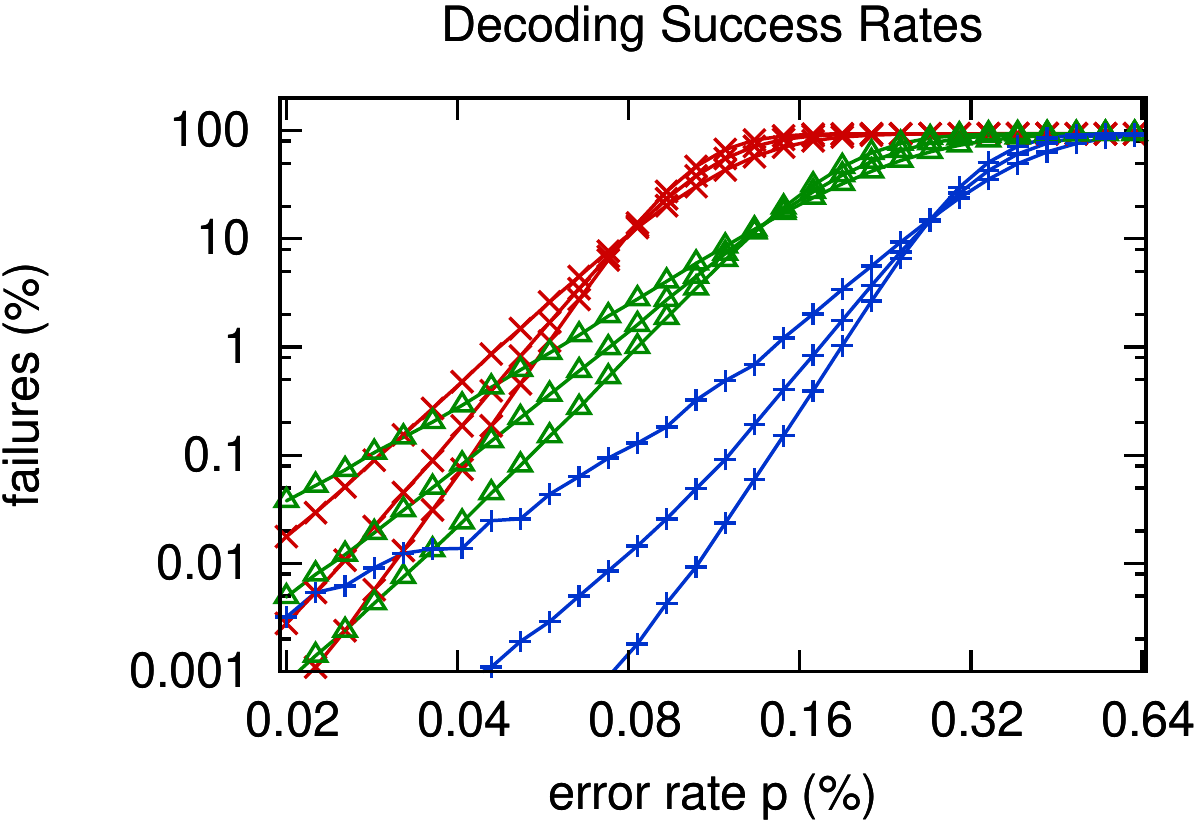,width=.375\textwidth}
   \label{subfig:log}
}
\fcaption{\label{fig:thCombined} (Color online) Comparisons of decoding success probabilities for the Full-LRU circuit and the Quick circuit with the Standard and HL decoder. (a) Linear scale on the left, (b) log scale on the right.}
\end{figure}

The threshold then decays as $\frac{\alpha}{1+\beta r}$ where $\alpha$ is a constant equal to the numeric value of the threshold in the regime without leakage. Assuming $\beta=3/4$ and choosing $\alpha$ such that the threshold at $r=0$ intersects the threshold of the Quick circuit with HL decoder, we plot the threshold of the ``idealized decoder'' in Fig.~\ref{fig:thresholds}~(a). A Standard decoder's threshold will not exceed that of the idealized one we just described, but a decoder that uses results from three-outcome measurements could in principle. Table~\ref{table:fitparameters} lists some extracted values of $\alpha$ and $\beta$ from the data in Fig.~\ref{fig:thresholds}~(a).

Fig.~\ref{fig:thresholds}~(b) shows that the threshold does not depend much on the relaxation rate $s$, likely because relaxation is much slower than the leakage reduction occuring in the circuits. This plot was obtained by recording thresholds for the Quick circuit with the HL decoder using the same technique as for Fig.~\ref{fig:thresholds}.

In our last set of simulations, we chose $r = s = 1$ and recorded the failure rate of the decoders as a function of $p$. Recall that $p_\uparrow = r p$ and $p_\downarrow = s p$. In Fig.~\ref{fig:thCombined} (a) we show failure rates for the Full-LRU circuit and the Quick circuit with both the Standard and HL Decoder. The thresholds of each circuit/decoder are visible as crossing points of the curves. The Full-LRU circuit has the lowest threshold, followed by the Quick circuit with the Standard decoder and the Quick circuit with the HL decoder. The success rate for the Partial-LRU circuit is not plotted because it was indistinguishable from the Quick circuit.

The same plot on the log scale in Fig.~\ref{fig:thCombined} (b) shows that when $p$ is well below threshold, the success rate of the Full-LRU circuit improves faster than for the Quick circuit with Standard decoder, which suggests that for low enough physical error rates the Full-LRU circuit will be better than the other circuits. We have a sufficient number of samples to observe at least $100$ failures at each data point shown in the figure. Table~\ref{table:fitparameters} presents the degree of error suppression $\gamma$ for each circuit based on fitting the logarithm of the logical error rate. For the HL decoder, we give the fit for each distance $d=7$, $9$, and $11$ since we appear to be observing the effects of higher order terms in the logical error rate and have not reached a sufficiently low physical error rate to capture the lowest order term in the polynomial. Since numeric simulations at extremely low physical error rates are inefficient and may require different techniques \cite{bravyi13}, we leave further exploration of this question for future work.

\begin{table}[t]
\tcaption{Threshold $\sim \frac{\alpha}{1+\beta r}$ and sub-threshold logical error rate $\sim Ap^{\gamma d}$}
\centerline{\footnotesize\smalllineskip
\begin{tabular}{r c c c}\\
\hline
{} & $\alpha$ & $\beta$ & $\gamma$ \\
\hline
{idealized} & $0.65$ & $3/4$ & n.a. \\
{Full-LRU} & $0.22$ & $1.72$ & $~0.67$\\
{Quick} & $0.65$ & $3.59$ & $~0.45$ \\
{Partial-LRU} & $0.55$ & $2.55$ & $~0.46$ \\
{Quick (HL)} & $0.62$ & $1.23$ & $~\{0.45,0.52,0.64\}$ \\
{Partial-LRU (HL)} & $0.55$ & $0.92$ & $~\{0.48,0.53,0.65\}$ \\
\hline\\
\end{tabular}}
\label{table:fitparameters}
\end{table} 
\section{Conclusion}\label{sec:conclusion}
\noindent
We provide a systematic study of error correction techniques that address leakage faults where qubit states leave the computational space. We developed a simple model that captures realistic elements of leakage phenomena. Using this model we studied the performance of various circuits that suppress leakage in the toric code, ranging from a simple circuit that only adds one $CNOT$ gate per syndrome extraction, to a scheme that reduces leakage after the application of every gate. To accurately compare the performance of these circuits, we considered two scenarios, one in which measurements distinguish leaked qubits from qubits in the computational state, and another in which they do not. For each of these cases we designed an optimized decoding algorithm and recorded its success rate in Monte Carlo simulations. Our results show that the Partial-LRU and Quick circuits are effective at reducing leakage and reduce the accuracy threshold of the toric code by less than a factor of $4$ in a realistic scenario where leakage was added to depolarizing noise. We see a factor of $2$ threshold improvement over Standard decoders for the HL Decoder that uses information about leakage detection events. Finally, there is evidence that the Full-LRU, which reduces leakage after every gate, may achieve the lowest logical error rates of all of the circuits when the physical error rate is less than $2\times 10^{-4}$. Future work could explore the logical error rates of these circuits at higher distances and at error rates that are far below threshold, perhaps by applying the splitting method \cite{bravyi13}.

\nonumsection{Acknowledgements}
\noindent
We thank Sergey Bravyi, Ken Brown, Oliver Dial, David DiVincenzo, Easwar Magesen, Graeme Smith, and John Smolin for helpful discussions, Chris Lirakis and Mark Ritter for their support, and Karen Bard for making Blue Gene available to us. AWC and JMG acknowledge support from ARO under contract {\em W911NF-14-1-0124}.

\nonumsection{References}
\noindent

%\bibliographystyle{unsrt}
%\bibliography{leakage}

\end{document}